\documentclass[aps,prl,10pt,twocolumn,showpacs,amsfonts,amssymb,amsmath]{revtex4-1}
\usepackage{amsmath}    
\usepackage{amsfonts}   
\usepackage{graphicx}   
\usepackage{subfigure}  
\usepackage{bm}         
\begin{document}
%
%
%
%
\title{Exact solution of bond percolation on small arbitrary graphs}
\author{Antoine~\surname{Allard}}
\author{Laurent~\surname{H\'ebert-Dufresne}}
\author{Pierre-Andr\'e~\surname{No\"{e}l}}
\author{Vincent~\surname{Marceau}}
\author{Louis~J.~\surname{Dub\'e}}
\affiliation{D\'epartement de physique, de g\'enie physique, et d'optique, Universit\'e Laval, Qu\'ebec (Qc), Canada G1V 0A6}
\date{\today}
\begin{abstract}
 We introduce a set of iterative equations that exactly solves the size distribution of components on small arbitrary graphs after the random removal of edges. We also demonstrate how these equations can be used to predict the distribution of the node partitions (i.e., the constrained distribution of the size of each component) in undirected graphs. Besides opening the way to the theoretical prediction of percolation on arbitrary graphs of large but finite size, we show how our results find application in graph theory, epidemiology, percolation and fragmentation theory.
\end{abstract}
\maketitle
%
%
%
%
%
\paragraph{Introduction.---}
%
Percolation on graphs is the study of the behavior of components in graphs whose nodes/edges are removed randomly. It has received a lot of attention recently for its various applications in many disciplines. Among them, let us mention the modeling of epidemic propagation \cite{salkeld10_pnas,davis08_nature,meyers07_bams,bansal06_plosmed} where the size distribution of components corresponds to the outbreak size distribution. The same distribution is also related to the size and composition of fragments in nuclear multifragmentation \cite{desesquelles11_plb,davila07_physA,paech07_prc,trautmann05_npa}. Finally, the study of the percolation threshold allows an assessment of the robustness (or reliability) of real networks to the failure of their nodes or edges \cite{buldyrevr10_nature,gallos05_prl,albert00_nature,callaway00_prl,kelmans65_ait,gilbert59_ams}. Alongside the intrinsic theoretical interest, this type of studies has triggered the development of increasingly realistic and complex models (see \cite{newman10_networks,cohen10_complex_networks,dorogovtsev08_rmp}, and references therein).

In the quest for ever more realistic models, a promising idea is to consider real networks not at the level of the nodes, but at higher levels of organization such as motifs, subgraphs or communities. It has recently been proposed that this perspective could help unify and explain many of the universal properties found in real networks \cite{hebertdufresne11_prl}. From the modeling perspective, using motifs as the fundamental building blocks of graphs has allowed to relax some of the limiting assumptions behind existing bond percolation models \cite{allard12a,karrer10_pre,gleeson09_pre,newman03b_pre}. This has effectively extended the class of models for which exact results can be obtained.

The advantages gained come at a price however: one needs to solve beforehand the distribution of the size of components in these motifs. While this can be done systematically for Erd\H{o}s-R\'enyi subgraphs (or cliques) \cite{gilbert59_ams,newman03b_pre}, the general problem must be solved by hand on a case-by-case basis. Hence, one is either limited to very small graphs or one has to rely on numerical simulations.

In this Letter, we introduce a set of iterative equations that exactly compute the size distribution of components in a multitype version of Erd\H{o}s-R\'enyi graphs. In the case where a different type is assigned to each node of a graph, the equations produce this distribution for any small arbitrary graphs, defined by an asymmetric nonnegative adjacency matrix, after the random removal of its edges. These equations therefore provide a systematic way to compute the required distributions in the motif-based bond percolation models mentioned previously.

We further show how these equations naturally offer a method to count the number of labeled multitype graphs with a given number of nodes and edges. We also explain how they can be used to study bond percolation on periodic infinite lattices. We finally demonstrate how these equations allow the exact calculation of the constrained distribution of the size of each component, or the partition of nodes, in undirected graphs. It is suggested that this provides a null model for fragmentation processes.
%
%
%
%
\paragraph{Percolation on multitype random graphs.---}
%
Let us consider a multitype generalization of the random graph model $\mathcal{G}_{n,p}$ \cite{gilbert59_ams}. These graphs are composed of $n$ nodes and an edge exists between any two nodes with probability $p$ regardless of other potential edges. We generalize this model by labelling nodes using $M$ types such that a graph of size $\bm{n}\equiv(n_1,\ldots,n_M)^\mathsf{T}$ is composed of $n_i$ type-$i$ nodes (with $i=1,\ldots,M$). Directed edges from type-$i$ nodes to type-$j$ nodes (noted $i \rightarrow j$) exist with probability $p_{ij}$ independently of one another. $p_{ij}$ need not be equal to $p_{ji}$ since edges are directed. Results statistically equivalent to undirected graphs --- where an undirected edge exists with a given probability --- are obtained in the symmetric case ($p_{ij}=p_{ji}$). While the usefulness of this multitype generalization will become manifest in the next section, multitype Erd\H{o}s-R\'enyi graphs can be used as a first approximation of structures in which correlations exist in the way nodes are connected (with an appropriate choice of $p_{ij}$).

We define $Q_i(\bm{l}|\bm{n})$ as the probability that the component reached from a type-$i$ node contains $\bm{l}\equiv(l_1,\ldots,l_M)^\mathsf{T}$ nodes (including the initial type-$i$ node) given that the graph contains $\bm{n}$ nodes. Because of the presence of directed edges, we extend the definition of a component to all nodes \textit{accessible} from a given node. Thus, node $A$ being in the component reached from node $B$ does not imply that the converse is true. In the same spirit of \cite{gilbert59_ams,newman03b_pre}, we now derive two recurrence equations allowing for the explicit calculation of $Q_i(\bm{l}|\bm{n})$.

\begin{figure*}[htb]
 \begin{center}
 \subfigure[]{\label{fig:allard_fig_1a} \includegraphics[width = 0.4\textwidth]{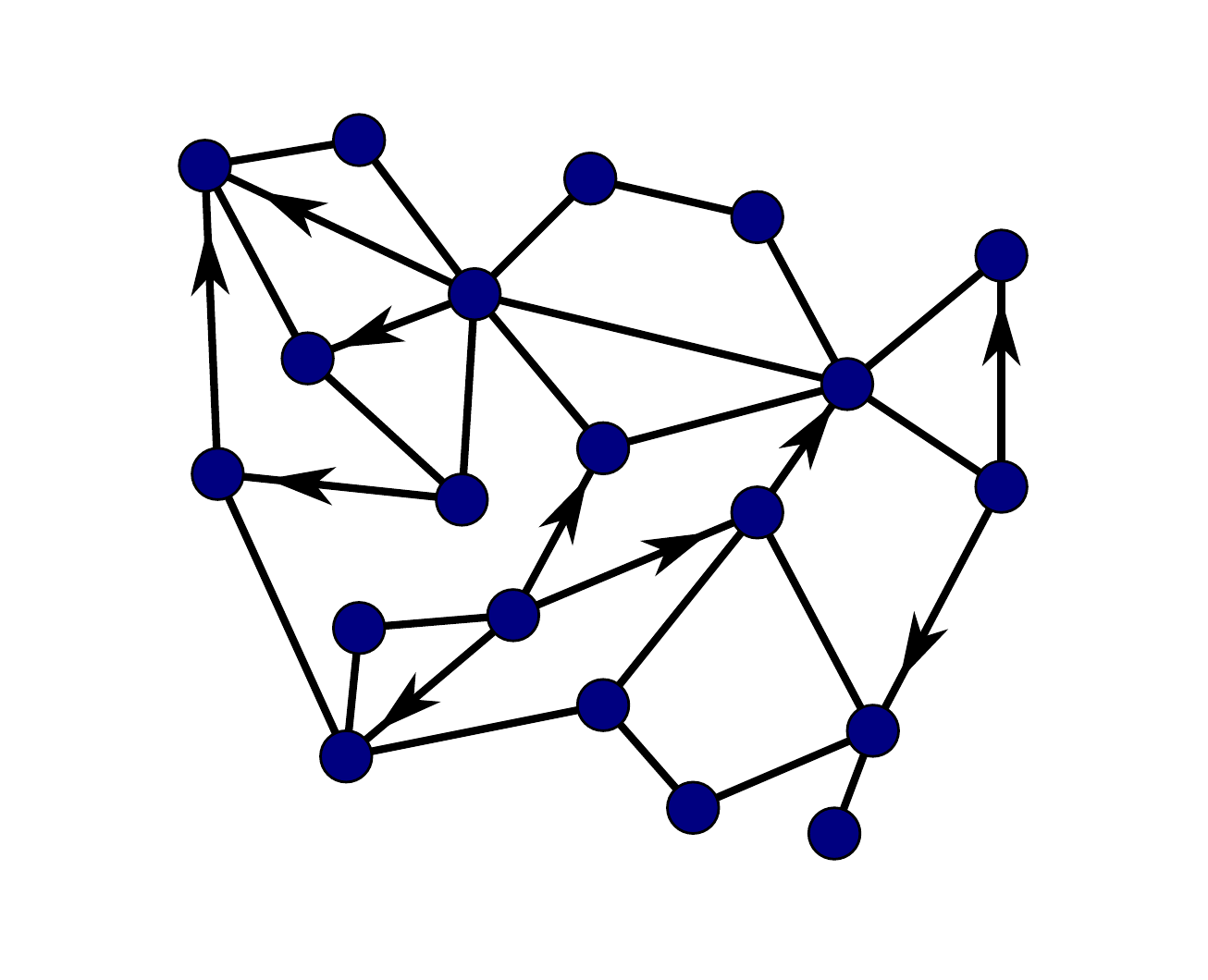}} \hspace{0.1\textwidth}
 \subfigure[]{\label{fig:allard_fig_1b} \includegraphics[width = 0.4\textwidth]{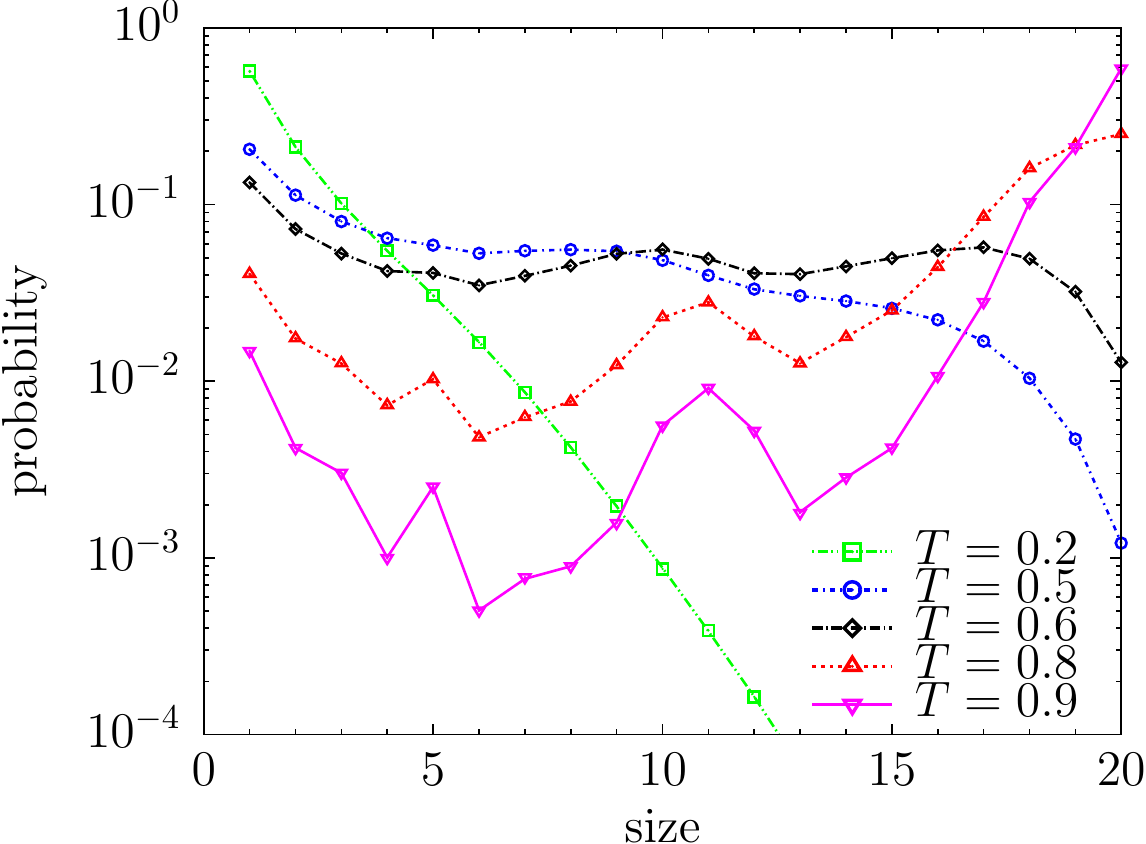}}
  \caption{\label{fig:allard_fig_1}(color online) (a) Example of an arbitrary graph of 20 nodes. Although not shown here, graphs with multiple edges could as easily be considered by our method. (b) Distribution of the size of the component reached from a randomly chosen node in the graph shown in (a) where edges have been removed with probability $1-T$. Lines were obtained using eqs.~(\ref{eq:gnp_2})--(\ref{eq:gnp_4}), and symbols were obtained by performing $10^7$ simulations. The distributions are discrete; lines have been added to guide the eye.}
 \end{center}
\end{figure*}

Let us consider a graph of size $\bm{n}$, a component of size $\bm{l}$, and an initial node of type $i$. We note $\delta_{ij}$ the Kronecker delta. Among the $n_j-\delta_{ij}$ nodes of type $j$ (excluding the initial node of type $i$), there are ${{n_j-\delta_{ij}} \choose {l_j-\delta_{ij}}}$ ways to choose the $l_j$ type-$j$ nodes that are part of the component. These $l_j$ type-$j$ nodes will not lead to any of the remaining $\bm{n}-\bm{l}$ nodes that are not part of the component with probability $\prod_{k}q_{jk}^{l_j(n_k-l_k)}$, where \mbox{$q_{jk}\equiv1-p_{jk}$}. In other words, this is the probability that no directed edge $j \rightarrow k$ exists between the $l_j$ type-$j$ nodes in the component and the $\bm{n}-\bm{l}$ nodes of every other types excluded from it. Repeating this procedure for the other types of nodes in the component, together with the observation that the $\bm{l}$ nodes form a component with probability $Q_i(\bm{l}|\bm{l})$, we have that
\begin{equation} \label{eq:gnp_1}
 Q_i(\bm{l}|\bm{n}) = Q_i(\bm{l}|\bm{l}) \prod_{jk} {{n_j-\delta_{ij}} \choose {l_j-\delta_{ij}}} q_{jk}^{l_j(n_k-l_k)} \ .
\end{equation}
That is, by knowing the probability of finding a component of size $\bm{l}$ in a graph of size $\bm{l}$, eq.~(\ref{eq:gnp_1}) computes the probability of finding a component of size $\bm{l}$ in a graph of size $\bm{n}$ (with $n_j \geq l_j$ for all $j$). Finally, to obtain $Q_i(\bm{l}|\bm{l})$, we note that the distribution $\{Q_i(\bm{m}|\bm{l})\}$ must be normalized for a given graph size $\bm{l}$, hence
\begin{equation} \label{eq:gnp_2}
 Q_{i}(\bm{l}|\bm{l}) = 1 - \sum_{\bm{m}<\bm{l}} Q_{i}(\bm{m}|\bm{l}) \ .
\end{equation}
The sum covers every possible instances of $\bm{m}$ such that $m_j \leq l_j$ for all $j$ but excludes the case where $m_j=l_j$ for each $j$. Starting with the initial condition $Q_{i}(\bm{\delta_i}|\bm{\delta_i}) = 1$, where $\bm{\delta_i} \equiv (\delta_{i1},\ldots,\delta_{iM})^\mathsf{T}$, we can therefore calculate every coefficient $Q_{i}(\bm{l}|\bm{n})$ using eqs.~(\ref{eq:gnp_1})--(\ref{eq:gnp_2}) iteratively. In other words, from a graph made out of a single node, eqs.~(\ref{eq:gnp_1})--(\ref{eq:gnp_2}) extend the graph to the desired size, and keep track of the component size distribution along the way to build the final distribution $Q_i(\bm{l}|\bm{n})$. A simple example of such a calculation is given in the Appendix. Setting $M=1$, we retrieve the recurrence equations presented in \cite{gilbert59_ams,newman03b_pre}. Also, a similar approach has been used to analyse the reliablity of communication networks \cite{kelmans65_ait}.

Using the identity $1=\prod_{j,k} (p_{jk}+q_{jk})^{n_j(n_k-\delta_{jk})}$ in eq.~(\ref{eq:gnp_2}), where $n_j(n_k-\delta_{jk})$ is the maximal number of $j \rightarrow k$ edges in the graph, the iteration of eqs.~(\ref{eq:gnp_1})--(\ref{eq:gnp_2}) yields polynomials whose coefficients have a direct combinatorial interpretation. Indeed, the coefficient in front of $\prod_{j,k}p_{jk}^{a_{jk}}q_{jk}^{b_{jk}}$ in $Q_{i}(\bm{l}|\bm{n})$ is simply the number of distinct ways to reach a component of size $\bm{l}$ in a graph of size $\bm{n}$ from a type-$i$ node using $a_{jk}$ existing and $b_{jk}$ absent $j \rightarrow k$ edges, respectively. The sum $a_{jk}+b_{jk}$ may be different than $n_{j}(n_k-\delta_{jk})$ as the existence of some edges may be irrelevant to the component. Of particuliar interest is the symmetric case where $p_{jk}=p_{kj}$ for all $j$ and $k$ for which $Q_{i}(\bm{n}|\bm{n})$ is independent of $i$ and whose coefficients are the number of connected labeled graphs of size $\bm{n}$. Hence, we see that eqs.~(\ref{eq:gnp_1})--(\ref{eq:gnp_2}) offer an alternative method to enumerate the number of graphs with a given number of edges and labeled nodes of different types.
%
%
%
%
%
\paragraph{Percolation on arbitrary graphs.---}
%
By considering that each node of an arbitrary graph belongs to its own type, eqs.~(\ref{eq:gnp_1})--(\ref{eq:gnp_2}) can exactly predict the outcome of a bond percolation process that has occurred on it. Predicting the outcome here is as precise as knowing the identity of the nodes that have been reached and of the ones that have not. To illustrate this point, let us consider the simplest case where an edge is to be kept with the same probability $T$ during the percolation process \footnote{The generality of eqs.~(\ref{eq:gnp_1})--(\ref{eq:gnp_2}) naturally allows for the use of various cases of type-dependent probabilities of existence of edges, all the way to the most general case where one specific probability is assigned for each direction of each edge.}. The probability $p_{jk}$ for the edge $j \rightarrow k$ to exist then becomes $p_{jk} = 1-(1-T)^{A_{jk}}$, where $A_{jk}\in\mathbb{N}$ is an element of the adjacency matrix $\mathbf{A}$ corresponding to the number of directed edges from node $j$ to node $k$. Arbitrary graphs with directed edges and multiple edges can thus be considered with our method.

Because each node belongs to its own type, the elements of the vectors $\bm{l}$, $\bm{m}$ and $\bm{n}$ indicate whether each node is present (value $1$) or not (value $0$). This allows us to write eq.~(\ref{eq:gnp_1}) in a simpler and more compact form
\begin{equation} \label{eq:gnp_3}
 Q_i(\bm{l}|\bm{n}) = Q_i(\bm{l}|\bm{l}) (1-T)^{\bm{l}^\mathsf{T}\mathbf{A}\bar{\bm{l}}} \ ,
\end{equation}
where the elements of $\bar{\bm{l}}$ are defined such that $l_j + \bar{l}_j = n_j$ for all $j$. That is, $\bm{l}^\mathsf{T}\mathbf{A}\bar{\bm{l}}$ is the number of outgoing edges that must not exist for the component of size $\bm{l}$ to be isolated from the rest of the graph. Using eqs.~(\ref{eq:gnp_2})--(\ref{eq:gnp_3}) together with the initial condition $Q_{i}(\bm{\delta_i}|\bm{\delta_i}) = 1$, we can calculate the exact probability of \textit{each individual outcome} of a percolation process on an arbitrary graph defined by its adjacency matrix $\mathbf{A}$.

To support this claim, fig.~\ref{fig:allard_fig_1b} compares the predictions of eqs.~(\ref{eq:gnp_2})--(\ref{eq:gnp_3}) with the results of numerical simulations of bond percolation on the graph shown in fig.~\ref{fig:allard_fig_1a}. To lighten the presentation of the results, fig.~\ref{fig:allard_fig_1b} shows the probability $q_k$ of finding a component of size $k$ --- regardless of the identity of the nodes --- from a randomly chosen node. This quantity is computed using
\begin{equation} \label{eq:gnp_4}
 q_k = \frac{1}{M} \sum_{i} \sum_{\bm{l}} Q_i(\bm{l}|\bm{n}) \delta\left( {\textstyle \sum_{j} l_j - k} \right)
\end{equation}
where $\delta( \cdots )$ is the delta function. We observe an excellent agreement between our theoretical predictions and the numerical results. Although for such a small network no precise percolation threshold can be defined, it is however clear that a qualitative change toward a ``giant component'' is initiated for $T \sim$ 0.5--0.6. Also, the irregular shape of the distribution for some values of $T$ highlights how $Q_i(\bm{l}|\bm{n})$ can depend on the precise structure of the graph. This advocates for the importance of developing methods that consider explicitly the structure of the graphs (i.e., the adjacency matrix).

Since eqs.~(\ref{eq:gnp_1})--(\ref{eq:gnp_3}) consider every possible outcomes of the percolation process, their predictions are exact. However, the calculational burden (e.g., required memory, number of operations) increases very quickly with the number of node types $M$. In the case of arbitrary graphs, it grows exponentially with the number of nodes. Thus, although eqs.~(\ref{eq:gnp_1})--(\ref{eq:gnp_3}) are in principle valid for graphs of any size, their use becomes cumbersome for large graphs. With our present computer facilities, a straightforward implementation of eqs.~(\ref{eq:gnp_2})--(\ref{eq:gnp_3}) have been able to handle graphs of size of the order of 25. A wiser implementation could certainly push this limit somewhat further. When dealing with a given graph, specific features of its structure may however be used to reduce the numerical effort. For instance, the distributions for different modules could be solved separately, and then recombined to obtain $Q_i(\bm{l}|\bm{n})$ for the whole graph. Quantum computation of the sort described in \cite{gaitan12_prl} may also be a solution for larger graphs.

Despite these limitations, our method compares favorably with an \textit{exact enumeration} method where a computer program explicitly considers each possible edge configuration, and then computes the component size distribution from them. Firstly, the computational demands of this approach scales exponentially with the number of edges $L$, whereas our approach scales exponentially with the number of nodes. The performance of our method should therefore be comparable to direct enumeration for sparse graphs, and should rapidly surpass it for denser graphs. Secondly, our method yields analytical solutions (i.e., polynomial in $p_{jk}$) valid for any value of $p_{jk}$.

Equations~(\ref{eq:gnp_2})--(\ref{eq:gnp_3}) can also be used to compute the bond percolation threshold of infinite periodic lattices. By virtue of the \textit{triangle-triangle} transformation \cite{scullard08_prl}, the percolation threshold is the root of a polynomial related to the connectivity of the basic cell of the lattice, which is a combination of the coefficients of $Q_i(\bm{l}|\bm{n})$. Thus our approach offers a systematic and exact way to compute this polynomial for complicated basic cells. The Appendix provides an example. Furthermore, our approach offers a systematic way to obtain the renormalisation-group transformation to estimate the scaling exponents and the bond percolation threshold of infinite lattices (see for instance eq.~(3.4) in \cite{reynolds80_prb}).

Finally, eqs.~(\ref{eq:gnp_1}) and (\ref{eq:gnp_3}) can be combined to compute $Q_i(\bm{l}|\bm{n})$ for graphs where nodes of different types interact through an arbitrary configuration of edges (see \cite{allard12a} for an explicit example). This allows to generate a wide range of realistic subgraphs (or motifs) found for instance in social networks, and to include them in motif-based bond percolation models \cite{allard12a,karrer10_pre}. As the component size distribution is closely related to the outbreak size distribution, our approach allows to study the spread of infectious diseases in more realistic urban settings \cite{meyers05_jtb}.
%
%
%
%
%
\paragraph{Predicting node partition distribution.---}
%
We can also use eqs.~(\ref{eq:gnp_1})--(\ref{eq:gnp_3}) to calculate the distribution of the number of components (and their size) found in an \textit{undirected} graph after the removal of a fraction of its edges. We restrict ourselves to undirected graphs because only in undirected graphs are components uniquely defined. That is, two nodes will be found in the same component with one unique probability regardless of the starting node. We illustrate how to perform the calculation using the $\mathcal{G}_{n,p}$ model. It should nevertheless be clear that equations for undirected multitype random graphs ($p_{ij}=p_{ji}$ for all $i$ and $j$) and undirected arbitrary graphs ($\mathbf{A}=\mathbf{A}^\mathsf{T}$) can be derived in a similar manner.
\begin{figure}[tb]
 \includegraphics[width = 0.4\textwidth]{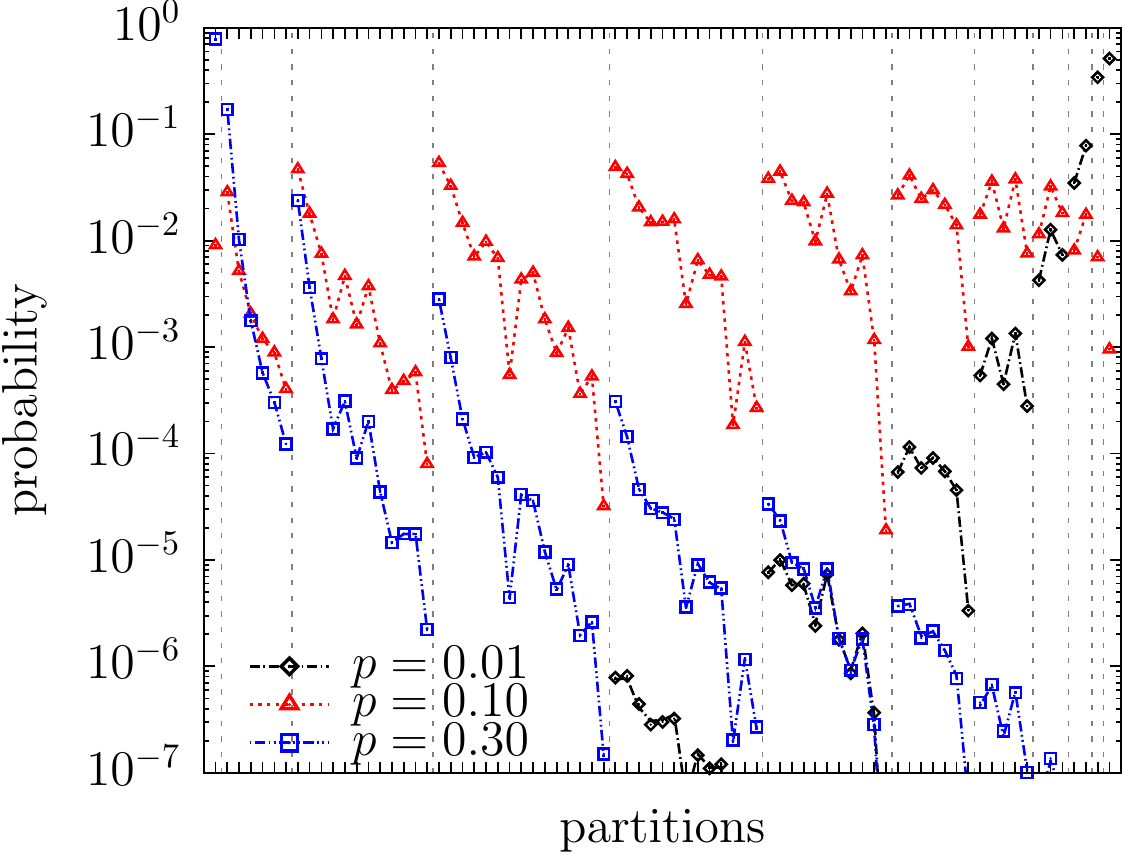}
 \caption{\label{fig:allard_fig_2}(color online) Probability of finding each node partition in $\mathcal{G}_{n,p}$ with $n=12$ and for various values of $p$. Lines were obtained using eq.~(\ref{eq:pr_1}), and symbols were obtained by performing over $5\times10^8$ simulations. The $|\mathcal{P}_{12}|=77$ integer partitions of 12 are displayed in an increasing order of the number of components, i.e., from the node partition where there is only one component (noted $\{12\}$) on the left to the case where there are 12 components of one single node (noted $\{1,1,1,1,1,1,1,1,1,1,1,1\}$) on the right. Linebreaks and vertical grey lines indicate where the number of components changes. Node partitions with the same number of components are displayed in an decreasing order of the largest components sizes (e.g., $\{8,2,2\}$ before $\{7,4,1\}$ before $\{7,3,2\}$).}
\end{figure}

Let us calculate the probability for a random graph composed of $n$ nodes to be split into $k$ components of size $\bm{r} \equiv (r_1$, $r_2$, \ldots, $r_k$) with $\sum_{i} r_i = n$. Each component $l$ will be connected, and therefore be a component, with probability $Q(r_l|r_l)$. We have dropped the subscript in $Q(r_l|r_l)$ because the probability to find one \textit{single} connected component in an undirected graph is the same regardless of the starting node. These components will be isolated from one another if none of the possible edges between nodes of different components exist. For the whole graph, this happens with probability $\prod_{i<j}(1-p)^{r_ir_j}$ where $p$ is the probability for an edge to exist between any two nodes.

The remaining step is to count the number of ways the $n$ nodes can be divided into $\bm{r}$ components, which we note $s_{\bm{r}}$. This number is in fact equal to the number of ways to put $n$ labeled objects into $k$ nonempty and \textit{unlabeled} containers of size $r_1$, $r_2$, \ldots, and $r_k$. First, let us point out that the number of ways to put $n$ labeled objects into $k$ \textit{labeled} containers whose sizes are given by $\bm{r}$ is simply the multinomial coefficient $n! / \prod_i r_i!$. To obtain $s_{\bm{r}}$, we just remove the redundant configurations due to containers of the same size. Hence, noting $d_m \equiv \sum_i \delta(m - r_i )$ the number of containers of size $m$, we get
\begin{equation}
 s_{\bm{r}} = \frac{n!}{\left(\prod_m d_m!\right) \left(\prod_i r_i!\right)} \ .
\end{equation}
Note that $s_{\bm{r}}$ is related to the \textit{Stirling number of the second kind} $\left\{ \begin{smallmatrix} n \\ k \end{smallmatrix}\right\}$ \cite{abramowitz_and_stegun} giving the number of ways to put $n$ labeled objects into $k$ nonempty and \textit{unlabeled} containers. Indeed $\left\{ \begin{smallmatrix} n \\ k \end{smallmatrix}\right\}$ is simply the sum of every $s_{\bm{r}}$ such that $\bm{r}$ has $k$ elements
\begin{equation}
 \left\{ \begin{matrix} n \\ k \end{matrix}\right\} = \sum_{\bm{r}\in\mathcal{P}_n} s_{\bm{r}} \delta\big( \dim(\bm{r}) - k \big) \ ,
\end{equation}
where $\mathcal{P}_n$ is the set of \textit{integer partition} of $n$, i.e., the set of decompositions of $n$ into a unordered sum of integers \cite{combinatorics_and_graph_theory}.

Combining these three contributions, we obtain the probability for a random graph composed of $n$ nodes to be split into $k$ components of size $\bm{r}$ to be
\begin{equation} \label{eq:pr_1}
 P(\bm{r}) = s_{\bm{r}} \prod_{l} Q(r_l|r_l) \prod_{i<j}(1-p)^{r_ir_j} \ .
\end{equation}

To validate eq.~(\ref{eq:pr_1}), fig.~\ref{fig:allard_fig_2} compares its predictions with the results obtained from numerical simulations of the $\mathcal{G}_{n,p}$ model for $n=12$ and for various values of $p$. Again, an excellent agreement between our theoretical predictions and the results of the numerical simulations is observed. Figure~\ref{fig:allard_fig_2} highlights the emergence of a ``giant'' component --- which occurs when $(n-1)p=1$ in the limit $n\rightarrow\infty$ --- as the distribution migrates toward partitions with fewer components with increasing $p$. It also shows that, for a same number of components, the partitions with larger components are more likely to occur in general. Although partly shown in fig.~\ref{fig:allard_fig_2}, this trend holds for all values of $p \in [0,1]$, and is due to the simple fact that there are more ways (i.e., possible configuration of edges) to build large connected components than small ones.

Our method can also serve as a null model for fragmentation processes. In fact, $P(\bm{r})$ is the probability for $n$ elements to be distributed among $\bm{r}$ fragments when bonds occur (or resist) randomly with probability $p$. Physical correlations can therefore be highlighted by comparing $P(\bm{r})$ with experimental data. Similar results for $\mathcal{G}_{n,m}$ \cite{erdos59_pmd}, where the number $m$ of edges (or energy) is fixed rather than its average value ${{n}\choose{2}}p$, have recently been used in the context of nuclear multifragmentation \cite{desesquelles11_plb}. As $\mathcal{G}_{n,p}$ and $\mathcal{G}_{n,m}$ can be seen as the ``canonical'' and ``micro-canonical'' version of Erd\H{o}s-R\'enyi random graphs, the results for $\mathcal{G}_{n,m}$ in \cite{desesquelles11_plb} can be reobtained using our equations (see the number of connected graphs at the end of the second section). Our method therefore emcompasses previous results, and fills the gap in contexts where the canonical approach is more relevant.
%
%
%
%
%
\paragraph{Conclusion.---}
%
We have introduced a set of iterative equations that computes the distribution of the size of the components in small random or arbitrary graphs. As directed and multiple edges can naturally be accounted for in the equations, our method is suitable for a wide range of arbitrary graphs. Because the equations consider systematically all possible outcomes of the bond percolation process, their predictions are exact. We have also demonstrated that they can be used to calculate the constrained distribution of the size of each component (i.e., node partition) for undirected small graphs. We have illustrated how these results find applications in various disciplines like graph theory, percolation theory, epidemiology and fragmentation theory. We believe that, despite the increasing unwieldiness of the calculation with the number of nodes, our results open the way to the theoretical prediction of bond percolation on large, but finite, arbitrary graphs.
%
%
%
%
%
%
\begin{acknowledgments}
  This work has been supported by the Canadian Institutes of Health Research (CIHR), the Natural Sciences and Engineering Research Council of Canada (NSERC), and Le Fonds de recherche du Qu\'ebec - Nature et technologies (FRQ-NT).
\end{acknowledgments}
%
%
%
%
%
\paragraph{Appendix: Explicit calculation of $Q_{i}(\bm{l}|\bm{n})$ and application.---}
%
We perform an explicit calculation of $Q_{i}(\bm{l}|\bm{n})$ to clarify the use of eqs.~(\ref{eq:gnp_1})--(\ref{eq:gnp_2}) and to illustrate some of our claims.

Let us consider a simple graph composed of 3 nodes of type 0, 1 and 2. The directed $i \rightarrow j$ edge exists with probability $p_{ij}$ and there are 6 possible directed edges. We note $q_{ij}=1-p_{ij}$. We take the node of type 0 as the starting node without loss of generality as the two other distributions can be obtained by permutation.

The calculation begins with the initial condition $Q_0(1,0,0|1,0,0) = 1$ stating the obvious fact that the probability of finding a component $\bm{l}=(1,0,0)$ in a graph of size $\bm{n}=(1,0,0)$ is 1. Equation~(\ref{eq:gnp_1}) provides the probability of finding the same component but in graphs respectively of size $(1,1,0)$, $(1,0,1)$ and $(1,1,1)$
\begin{align*}
 Q_0(1,0,0|1,1,0) & = q_{01} \\
 Q_0(1,0,0|1,0,1) & = q_{02} \\
 Q_0(1,0,0|1,1,1) & = q_{01}q_{02} \ .
\end{align*}
Using eq.~(\ref{eq:gnp_2}), we then compute the probability to find a component of size 2 in the graphs of size $(1,1,0)$, $(1,0,1)$
\begin{align*}
 Q_0(1,1,0|1,1,0) & = 1 - Q_0(1,0,0|1,1,0) = p_{01} \\
 Q_0(1,0,1|1,0,1) & = 1 - Q_0(1,0,0|1,0,1) = p_{02} \ .
\end{align*}
We use once more eq.~(\ref{eq:gnp_1}) to compute the probability of finding the same components but in a graph of size $\bm{n}=(1,1,1)$
\begin{align*}
 Q_0(1,1,0|1,1,1) & = p_{01}q_{12}q_{02} \\
 Q_0(1,0,1|1,1,1) & = p_{02}q_{21}q_{01} \ .
\end{align*}
Finally, the probability of reaching the whole graph of size $(1,1,1)$ is obtained with eq.~(\ref{eq:gnp_2})
\begin{align*}
 Q_0(1,1,1|1,1,1) & = 1 - q_{01}q_{02} - p_{01}q_{12}q_{02} - p_{02}q_{21}q_{01} \\
                  & = p_{01}p_{02} + p_{01}p_{12}q_{02} + p_{02}p_{21}q_{01} \ ,
\end{align*}
where we have used the identity
\begin{equation*}
 1 = (p_{01}+q_{01})(p_{02}+q_{02})(p_{12}+q_{12})(p_{21}+q_{21})
\end{equation*}
to obtain a polynomial with positive coefficients. As claimed previously, each term in this last polynomial can be interpreted as a path leading to the component $\bm{l}=(1,1,1)$, and its coefficient as the number of distinct realisations of such a path.

With these results, we show how to compute the percolation threshold $p_c$ for the infinite triangular lattice using the \textit{triangle-triangle} transformation \cite{scullard08_prl}. Setting $p_{ij}=p$ for all $i$ and $j$ in $Q_0(1,1,1|1,1,1)$ with $q=1-p$, we retrieve the probability for the three nodes to be connected in some way, that is $p^3 + 3 p^2q$. Remember that for undirected graphs, the probability of reaching the entire graph is independent of the starting node. The \textit{triangle-triangle} transformation then stipulates that $p_c$ is the lowest value in [0,1] for which this last probability is equal to the probability $q^3$ that none of the nodes are connected. Thus, $p_c$ satisfies
\begin{equation*}
 p_c^3 - 3p_c + 1 = 0 \ ,
\end{equation*}
whose only solution in [0,1] is $p_c=2 \sin \left(\pi/18\right)$, which is the exact value of the bond percolation threshold for the triangular lattice \cite{sykes64_JMathPhys}.

This simple example could have been solved without using eqs.~(\ref{eq:gnp_1})--(\ref{eq:gnp_2}). However, repeating this exercise for graphs of 4, 5 or 6 nodes should convince the reader that a systematic procedure, as provided by eqs.~(\ref{eq:gnp_1})--(\ref{eq:gnp_2}), quickly becomes necessary.
%
%
%
%
%
%
%
%
%
%
\end{document}